\documentclass[letter,twocolumn]{jpsj3}
\usepackage{graphicx,amsmath,amssymb,bm}

\allowdisplaybreaks 

\usepackage{color}
\definecolor{Green}{rgb}{0,0.7,0}
\newcommand{\e}{ {\rm e}}
\newcommand{\h}{ {\varepsilon}}
\newcommand{\Dk}{ {\Delta_{\bm{k}}}}
\newcommand{\Dq}{ {\Delta_{\bm{q}}}}

\def \bmk{{\bm{k}}}
\def \bmq{{\bm{q}}}
\def \bmG{{\bm{G}}/2}
\def \ET{{ $\alpha$-(BEDT-TTF)$_2$I$_3\; $}}

\begin{document}

\title{
Analysis of Dirac Point 
 in the Organic Conductor $\alpha$-(BEDT-TTF)$_2$I$_3$ 
}
\author{
Yoshikazu Suzumura 
}
\inst{
Department of Physics, Nagoya University, 
 Chikusa-ku, Nagoya 464-8602, Japan 
}

\recdate{March  \;\;\;, 2016}
\abst{
The  Dirac electron in the organic conductor \ET under pressure 
 is analyzed using a tight-binding model with 
   nearest-neighbor transfer energies and four molecules per unit cell.   
By noting that the Dirac point between the first and second  energy bands  emerges or merges  followed by the level crossing  at a time-reversal invariant momentum (TRIM), 
an effective Hamiltonian  is derived on the basis of these two 
 wave functions at the TRIM,  
 which have different parities  associated with  an inversion symmetry  around the inversion center. 
We demonstrate that   
 the Dirac point  is determined by  an intersection of  two kinds of lines 
   originating   from  the   Hamiltonian 
 described by  symmetric and antisymmetric functions around  the TRIM. 
The present method   quantitatively gives  a reasonable  location 
 of the  Dirac point 
 of $\alpha$-(BEDT-TTF)$_2$I$_3$ in a wide  pressure range. 
 }

\maketitle

It is well known that the organic conductor \ET exhibits the Dirac point 
 on the Fermi energy  under pressure.\cite{Katayama2006_JPSJ75} 
 By utilizing  the energy  band  with such a Dirac cone,
several studies  have been performed to explore the properties  
 of the Dirac electron.\cite{Kajita_JPSJ2014}
However, in contrast to graphene,\cite{Novoselov2005_Nature438,Ando2005_JPSJ74}
  it is not straightforward to obtain the Dirac point on the Brillouin zone
owing to  the accidental degeneracy at the Dirac point.\cite{Herring}

The existence of such a Dirac point is verified using a 
 product  of the parity at the time-reversal invariant momentum (TRIM),
\cite{Fu2007_PRB76,Mori2013_JPSJ,Piechon2013_JPSJ,Kariyado2013_PRB}  
 and  the merging or emergence of a pair of Dirac points occurs at the TRIM 
 followed by level crossing.\cite{Piechon2013_JPSJ} 
The node of the component of the wave function\cite{Kobayashi2013_JPSJ} 
 connects the Dirac point with the TRIM resulting in the Berry phase.
\cite{Piechon2013_Berry} 
Although  the correspondence  between the Dirac point and the TRIM has been clarified 
  in these works, it is not yet successful to determine  the location of the Dirac point
 except for the numerical diagonalization or semi-analytical calculation.
\cite{Suzumura2013_JPSJ} 
Thus, further study of the role  of  the TRIM is needed 
 to elucidate the mechanism of the formation  of the Dirac point.
In this present paper, a method of finding  the Dirac point is demonstrated 
   using  an effective Hamiltonian
 based on the parity of the wave  function  at the TRIM.

\begin{figure}
  \centering
\includegraphics[width=6cm]{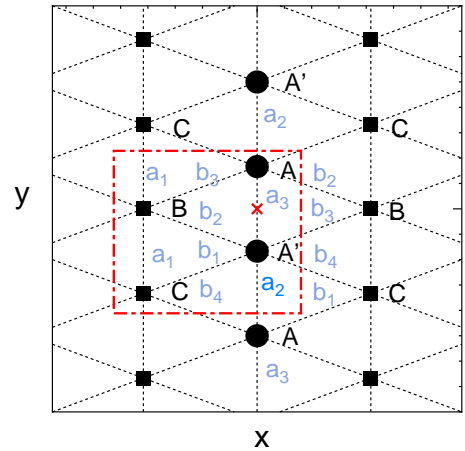}   
  \caption{(Color online)
Structure of $\alpha$-(BEDT-TTF)$_2$I$_3$ on the x-y plane 
 with four molecules   A, A', B, and C per unit cell (dot-dashed square). 
The bond (dotted line) shows nearest-neighbor transfer energies,  $ a_1 , \cdots, b_4$. 
 The cross denotes an inversion center located at  the middle  of  A and A' sites.\cite{Mori1984_CL}.
}
\label{fig:structure}
\end{figure}

Figure \ref{fig:structure} shows  the structure 
of $\alpha$-(BEDT-TTF)$_2$I$_3$ with four molecules A, A', B, and C 
 in the unit cell (the dot-dashed square),
    which gives  a tight-binding model  
      with seven  kinds of nearest-neighbor transfer energies 
 shown by a bond, 
        $a_1, \cdots, b_4$.\cite{Mori1984_CL}    
There is an inversion symmetry around  the sites of   
 B,  C, and  the middle of  A  and A'.
 The tight-binding model of Fig.~\ref{fig:structure} 
 is written as 
\begin{eqnarray}
\hat{H}_0 = \sum_{\bm{l},\bm{l}'} \sum_{\alpha,\beta}
   |\bm{l'}\beta > 
t_{\alpha, \beta ; \bm{l},\bm{l'}}
 <\bm{l}\alpha |  \; ,
\label{H_0}
\end{eqnarray}
where $|\bm{l}\alpha>$ denotes a state vector on the molecular site, and   
   $t_{\alpha,\beta;\bm{l},\bm{l}'}$ is 
  the transfer energy between nearest-neighbor  molecular sites. 
 The quantities $\alpha$ and  $\beta$   denote 
  the sites A, A', B, and C, and $\bm{l}$ and $\bm{l}'$ are 
 the position vectors of the cell forming a square lattice with $N$ sites. 
 By using  the Fourier transform 
 $|\bm{l}\alpha> = N^{-1/2} \sum_{\bm{k}}
 \exp [ i \bm{k} \bm{l}] |\bm{k}\alpha>$
 with a wave vector  $\bm{k}=(k_x,k_y)$,  
Eq.~(\ref{H_0}) is rewritten as 
$
\hat{H}_0 = \sum_{\bmk} 
  |\Psi (\bmk)> H_S <\Psi (\bmk)|
 $, 
where 
$
|\Psi (\bmk)> = (
2^{-1/2}(|\bmk A> +|\bmk A'>), 
2^{-1/2}(|\bmk A> -|\bmk A'>), 
|\bmk B>, |\bmk C>
)^t
$ with $H_S$ being a 4 x 4 matrix Hamiltonian.

By using a unitary transformation,
  $H_S(\bmk)$ is rewritten 
   as a real Hamiltonian,\cite{Piechon2013_Berry}  
\begin{eqnarray}
  {H}(\bm{k})  
 &=& {P}(\bm{k})^{-1/2}   {H}_S(\bm{k})  {P}(\bm{k})^{1/2}
 \nonumber \\ 
 &=& 
\begin{pmatrix}
\h_{11} & \h_{12}& \h_{13}& \h_{14} \\
\h_{21} & \h_{22} & \h_{23}& \h_{24}\\
\h_{31} & \h_{32} & \h_{33} & \h_{34}\\
\h_{41} & \h_{42} & \h_{43} & \h_{44}
\end{pmatrix} \ , 
\label{eq:eq2}
\end{eqnarray}
where 
$\h_{11} = a_3+a_2 \cos k_y$, $\h_{22} = -\h_{11}$, 
$\h_{12} =  -  a_2 \sin k_y$,  
$\h_{13} = \sqrt{2}(b_2 + b_3) \cos (k_x/2)$,
$\h_{14} = 
           \sqrt{2} b_1 \cos (\frac{k_x+k_y}{2})
          +  \sqrt{2} b_4 \cos (\frac{k_x-k_y}{2})$,
$\h_{23} = \sqrt{2}(b_3 - b_2) \sin(k_x/2)$,
$\h_{24} = 
          - \sqrt{2} b_1 \sin (\frac{k_x+k_y}{2})
           + \sqrt{2} b_4 \sin (\frac{k_x-k_y}{2})$, and 
$\h_{34} =  2 a_1 \cos (k_y/2) $. 
 $\h_{33}=h_{44}=0$
 and   $\h_{ij}=\h_{ji}$.  
 ${P}(\bm{k})$ denotes a matrix for the transformation of the base 
 by  a $\pi$-rotation  around the inversion center, 
 which is taken at the middle  between A and  A'.
 The elements of $P(\bm{k})$ are given by 
$[P(\bm{k})]_{\gamma,\gamma'}
 =  p_{\gamma}(\bm{k}) \delta_{\gamma,\gamma'}$
 with  $p_{\gamma}(\bmk) = 1, -1$, $ \e^{-ik_x}$, and  $\e^{-i(k_x+k_y)}$
 for $ \gamma$ = 1, 2, 3, and 4, respectively. 
The equation for  $P(\bm{k})$ is written as    
\begin{eqnarray}
P(\bm{k})|u_\gamma> = p_{\gamma}(\bmk)|u_\gamma> \; ,
\label{parity:eigenvalue}
\end{eqnarray}
 which gives 
  $p_{\gamma}(\bmk) = 1, -1$, $ \e^{-ik_x}$, $\e^{-i(k_x+k_y)}$, 
 and $|u_\gamma> = $
$(1,0,0,0)^{t}$, 
$(0,1,0,0)^{t}$, 
$(0,0,1,0)^{t}$, 
$(0,0,0,1)^{t}$, 
respectively.
The TRIM is given by  $\bm{k}=\bm{G}/2$ 
  with  $\bm{G}$ being the reciprocal lattice vector, 
where  $\bm{G}/2$ is written as 
 $(k_x/\pi,k_y/\pi)$ = (0,0) ($\Gamma$ point),
 (1,0) (X point), (0,1) (Y point), and (1,1) (M point).

We mention  two factors for deriving the effective Hamiltonian.
 One is the parity of the wave function at the TRIM, 
 which is obtained by  $P(\bmG)$.\cite{Piechon2013_JPSJ} 
 Noting   $p_{\gamma}(\bm{G}/2) = \pm 1$ in Eq.~(\ref{parity:eigenvalue})
 and a relation 
\begin{eqnarray}
[P(\bm{G}/2),H(\bm{G}/2)]=0 \; ,
\label{eq:commutation} 
\end{eqnarray}
one obtains  
\begin{eqnarray}
P(\bm{G}/2) |j(\bm{G}/2)> &=& p_{Ej}(\bmG) |j(\bm{G}/2)> \; , 
   \\
H(\bm{G/2})|j(\bm{G}/2)> &=&  E_j(\bm{G}/2)|j(\bm{G}/2)> \; ,
\label{eq:eq6}
\end{eqnarray}
 where $p_{Ej}(\bmG) = \pm 1$,   and   $E_1 > E_2 > E_3 > E_4$. 
Since the  $+ (-)$ sign gives   the even (odd) parity for both 
$|u_{\gamma}>$ and  $|j(\bm{G}/2)>$,  
 the wave function 
$|j(\bm{G}/2)>= \sum_{\gamma}d_{j\gamma}(\bm{G}/2)|u_{\gamma}>$,
with the even (odd) parity, is described  by a linear combination of $|u_\gamma>$ with  even ( odd) parity. 
The other is the relation between the parity and the matrix element 
 of Eq.~(\ref{eq:eq2}).
When the parity of $|i(\bm{G}/2)>$ is different from that of  $|j(\bm{G}/2)>$,
 \begin{eqnarray}
&&<i(\bm{G}/2)|H(\bm{G}/2) |j(\bm{G}/2)> \nonumber \\
 && =\sum_{\gamma_1,\gamma_2}
  d_{i\gamma_1}(\bm{G}/2) \h_{\gamma_1,\gamma_2} d_{j\gamma_2}(\bm{G}/2)
 = 0 \;  ,
\label{eq:eq7}
\end{eqnarray}
leads to  
\begin{eqnarray}
\h_{\gamma_1,\gamma_2}(\bm{G}/2)     = 0 \; ,
\label{eq:eq8}
\end{eqnarray}
 for   $p_{\gamma_1}(\bm{G}/2)  p_{\gamma_2}(\bm{G}/2) = -1$. 
In general, $\h_{\gamma_1,\gamma_2}(\bm{G}/2)   \not= 0$  
 for  $p_{\gamma_1}(\bm{G}/2)  p_{\gamma_2}(\bm{G}/2) = 1$.  
Thus, for $p_{\gamma_1}(\bm{G}/2) p_{\gamma_2}(\bm{G}/2) = -1 (+1)$, 
 the matrix element  $\h_{\gamma_1,\gamma_2}(\bmk)$  
with $\bmk = \bmG + \bmq$ 
is antisymmetric (symmetric) as a function of $\bm{q}$ 
 owing to $E_j(\bmG+\bm{q}) = E_j(\bmG - \bm{q})$, i.e., the time-reversal symmetry. 
 In the case of  the $\Gamma$ point ($\bmG$=0),
 the antisymmetric ( symmetric) function with respect to $\bmk$ is given by  
   $\h_{12}(\bmk)$, $\h_{23}(\bmk)$, and  $\h_{24}(\bmk)$ (the others) in Eq.~(\ref{eq:eq2}).
When  $|i(\bm{G}/2)>$ and  $|j(\bm{G}/2)>$ have different parities, 
the quantity  $<i(\bm{G}/2)|H(\bmG + \bmq)|j(\bm{G}/2)>$
 becomes antisymmetric as a function of $\bm{q}$,  and then 
 \begin{eqnarray}
<i(\bm{G}/2)|H(\bmG + \bmq)|j(\bm{G}/2)> =  0 \; ,
\label{eq:eq9}
\end{eqnarray}
  gives a line passing through  $\bm{G}/2$ owing to $\h_{\gamma_1,\gamma_2}$ being real.
 
Now, we examine  
the Dirac point between 
 $E_1(\bmk)$ and $E_2(\bmk)$,  which emerges 
 at $\bmk = \bmG$ followed by  
 a level crossing of  $E_1(\bm{G}/2)$ and $E_2(\bm{G}/2)$
 with different parities. 
Taking  these  $|1>$ and $|2>$  
  as the unperturbed states  where 
  $E_j [\equiv E_j(\bm{G}/2)]$ and $|j> [\equiv |j(\bm{G}/2)>]$, 
 we calculate $E_1(\bmk)$ and $E_2(\bmk)$ 
 at an arbitrary $\bmk$  by the  perturbation of  $V_\bmk$ defined as     
 \begin{eqnarray}
 V_\bmk =H_\bmk - H(\bm{G}/2) \; ,
 \label{eq:eq10}
\end{eqnarray}
 and $H_\bmk = H(\bm{k})$.
 By extending our previous works,\cite{Katayama2009_EPJB57,Kobayashi2010_PRB84}
 in which 
the base is  taken at  the Dirac point 
 or the M point in the charge-ordered state, 
 the  effective Hamiltonian 
up to the second order 
is written as 
\begin{eqnarray}
H_{\rm eff}(\bm{k})= 
\begin{pmatrix}
  <1|\tilde{H}_\bmk|1> & <1|\tilde{V}_\bmk|2>  \\
 <2|\tilde{V}_\bmk|1> & <2| \tilde{H}_\bmk|2>
\end{pmatrix} \; , 
 \label{eq:eq11}
\end{eqnarray}
where  $\tilde{H}_\bmk = H(\bm{G}/2)+ \tilde{V}_\bmk$ and $\tilde{V}_\bmk$ is given by
\begin{eqnarray}
&& <i|\tilde{V}_\bmk|j>
   = <i|V_\bmk|j> 
+ \frac{1}{2} \sum_{n=3,4}    <i| V_{\bmk} |n> 
     \nonumber \\
&& \times \left(
  \frac{1}{E_i-E_n}+\frac{1}{E_j-E_n} \right)
  <n| V_\bmk|j>  \; . 
    \nonumber \\
 \label{eq:eq12}
\end{eqnarray}
  From  Eq.~(\ref{eq:eq11}), the energies   
 $\tilde{E}_1(\bmk)$ and $\tilde{E}_2(\bmk)$
   and the gap function  $\Dk$  are estimated as 
\begin{eqnarray}
 \tilde{E}_{j}(\bmk)& =& 
 \frac{<1|\tilde{H}_{\bmk}|1> + <2|\tilde{H}_{\bmk}|2>}{2}  -(-1)^{j}  \frac{\Dk}{2}  
 \; , \;\;\;\;\;\;
    \nonumber  \\                
 \label{eq:eq13}
 \\ 
 \Dk 
 &=& \tilde{E}_1(\bm{k}) -\tilde{E}_2(\bm{k}) 
 = \sqrt{f(\bm{k})^2 + g(\bm{k})^2} \; ,
  \label{eq:eq14}
\end{eqnarray}
where  
%
\begin{eqnarray}
 f(\bm{k}) & = &  2 <1|\tilde{H}_\bmk|2> \; ,
\label{eq:eq15}
   \\
 g(\bm{k}) & = & <1|\tilde{H}_\bmk|1>-<2|\tilde{H}_\bmk|2> \; .
\label{eq:eq16}
\end{eqnarray}
 The coupling  between two bands of 
 $<1|\tilde{H}_{k}|1>$ and  $<2|\tilde{H}_{k}|2>$ vanishes 
 on a line given by  $f(\bmk)=0$ [Eq.~(\ref{eq:eq9})],  
while  
   $ <1|\tilde{H}_{k}|1>$ becomes  equal to  $<2|\tilde{H}_{k}|2>$
  on a line of  $g(\bmk)=0$, as will be shown later.   
 The  Dirac point $\pm \bm{k}_0$  is obtained from 
\begin{eqnarray}
  \Delta_{\bm{k}_0}=  0  \; , 
 \label{eq:eq17}
\end{eqnarray} 
 which is equivalent to $f(\bm{k}_0) = 0$ and $g(\bm{k}_0) = 0$.   
The effective Hamiltonian [Eq.~(\ref{eq:eq11})] consisting  of  
 the antisymmetric $f(\bm{k})$ and  symmetric $g(\bm{k})$
 has a common feature with  that  of Ca$_3$PbO \cite{Kariyado2012}
 in which the Dirac point is obtained using  
the Ca-$d_{x^2-y^2}$  and  Pb-$p$ orbitals with different  symmetries.

We numerically examine the case of $\alpha$-(BEDT-TTF)$_2$I$_3$.
The transfer energy $t = a_1, \cdots, b_4$ (eV) 
 at the uniaxial pressure $P$ (kbar) is  estimated as\cite{Kondo_2005,Katayama2006_JPSJ75} 
\begin{eqnarray}
t(P) = t(0) (1+K_t P)  \; , 
 \label{eq:eq20}
\end{eqnarray} 
 where 
$a_1(0), \cdots, b_4(0)$ = 
$-0.028, -0.048, 0.020, 0.123, 0.140, 0.062, 0.025$,
 and the corresponding $K_t$ is given by 
 $0.089, 0.167, -0.025, 0, 0.011$, and 0.032, respectively.

The calculation of the  Dirac point in terms of  
  the states  of the TRIMs ($\bmG$ =  $\Gamma$, X, Y, and M) 
  is performed as follows.
 (i) At a given pressure, verify  the existence of the Dirac point using
   the  condition  
$p_{E1}(\Gamma) p_{E1}({\rm X}) p_{E1}({\rm Y}) p_{E1}({\rm M}) = - 1$.\cite{Piechon2013_JPSJ}
(ii) Vary  (increase and decrease) pressure to obtain the TRIM 
  at which  
 the level crossing of $E_1(\bmG)$ and $E_2(\bmG)$  with different parities 
 occurs. In this case, 
  the Dirac point moves to the respective  TRIM 
 owing to the  merging of a pair of Dirac points. 
(iii) Calculate the Dirac point   by substituting  the state of  the TRIM 
  into Eq.~(\ref{eq:eq17}). 
 One of these TRIMs is chosen depending on the location of the Dirac point.

 The parity of the wave function in the present case 
is found  as 
 [$(p_{E1}(\bmG), p_{E2}(\bmG)$] 
  = 
 $(+,-), (-,-), (+,-)$, and $(+,-)$ 
  at the $\Gamma$, X, Y, and M points respectively. 
Then the present method of Eq.~(\ref{eq:eq11}) may be applied  
  to the $\Gamma$, Y, and M points  owing to  
 $p_{E1}(\bm{G}/2) p_{E2}(\bm{G}/2) = -1$. 

\begin{figure}
  \centering
\includegraphics[width=4cm]{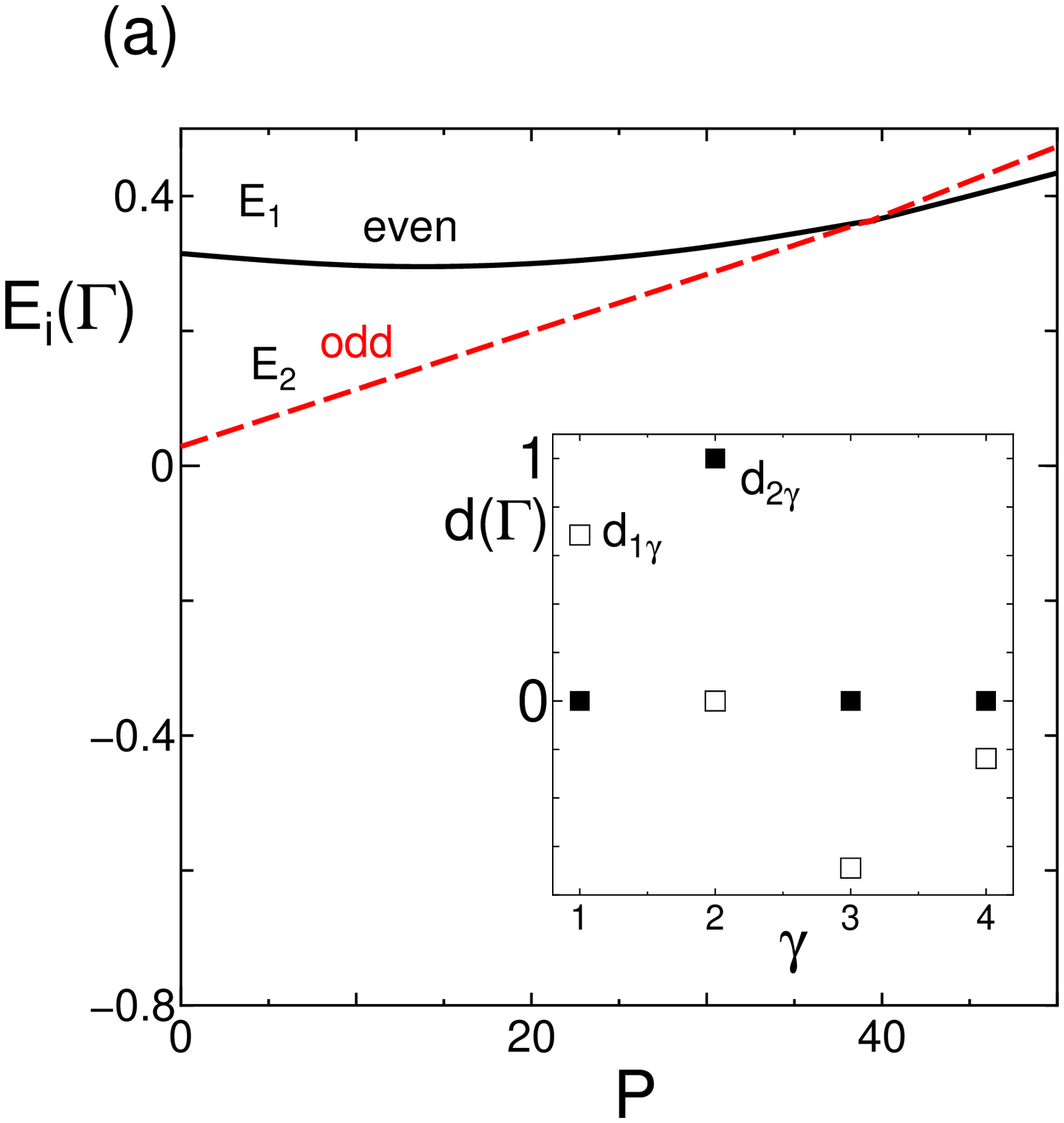}   
\includegraphics[width=4cm]{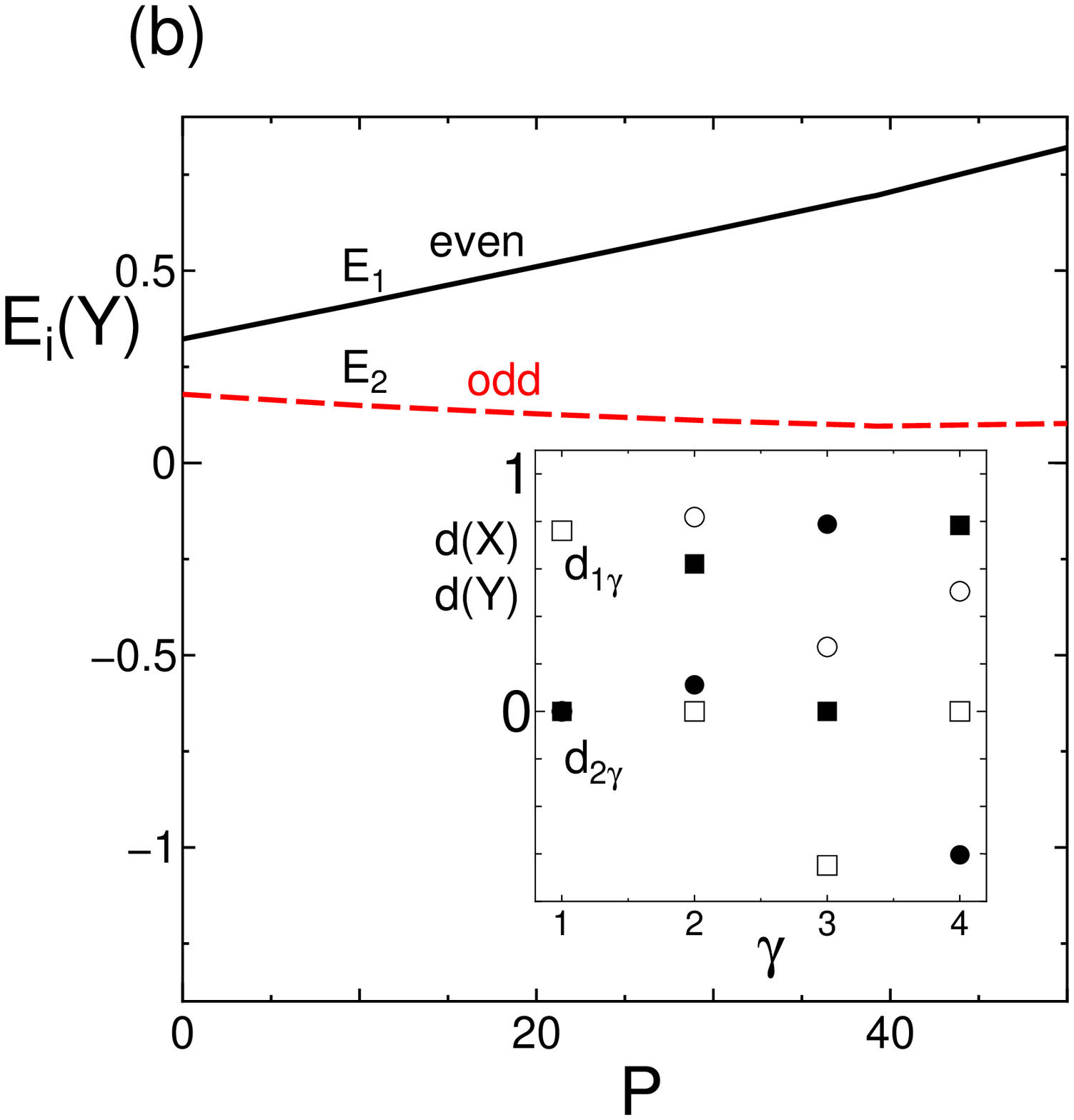}   
  \caption{(Color online)
Pressure ($P$) dependence of $E_j(\bm{G}/2)$ 
 ($j$=1 and 2) with  the even or odd parity
 where  $\bmG$ corresponds  to 
 the $\Gamma$ point (a) and  Y point (b).   
 The inset shows  components of $d_{j\gamma}(\bmG)$ for  
  $E_1$ (open square) and $E_2$ (closed square)
 at  $P$=38 kbar in (a),   
  and at $P$ =4 kbar in (b)  
 where the open and closed circles in (b) 
 denote $d_{j \gamma}(X)$ ($j$=1 and 2, respectively) at 4 kbar. 
}
\label{Fig2}
\end{figure}

In Figs.~\ref{Fig2}(a) and 2(b),
 the pressure dependence of $E_j(\bm{G}/2)$ ($j$=1 and 2) 
 is shown for the $\Gamma$ and Y points where  
 the level crossing  of $E_1(\bmk)$ and $E_2(\bmk)$ occurs   
  at the $\Gamma$ point for  $P=P_0$ (= 39.2 kbar),  
 and at the Y point for  $P$ = -11.8 kbar (from the extrapolation), respectively. 
The  Dirac point for  $P \sim P_0$  
is examined using the  
  wave function of the $\Gamma$ point, 
 while the wave function of the Y point  is applied to   
  the Dirac point being far away from the $\Gamma$ point.  
The inset denotes the corresponding $d_{j\gamma}(\Gamma)$ 
  at $P$ = 38 kbar (a) 
 and $d_{j\gamma}(Y)$  at $P$ = 4 kbar (b)
  [and $d_{j\gamma}(X)$ for comparison].
Each component has either  the even parity or the odd parity.
For the $\Gamma$ point,
 the wave functions of $E_1$, $E_3$, and $E_4$ ($E_2$)
  have the even (odd) parity.
 For the Y point, 
 the wave functions of $E_1$ and $E_3$, ($E_2$ and $E_4$)
  have the even (odd) parity.

\begin{figure}
  \centering
\includegraphics[width=4cm]{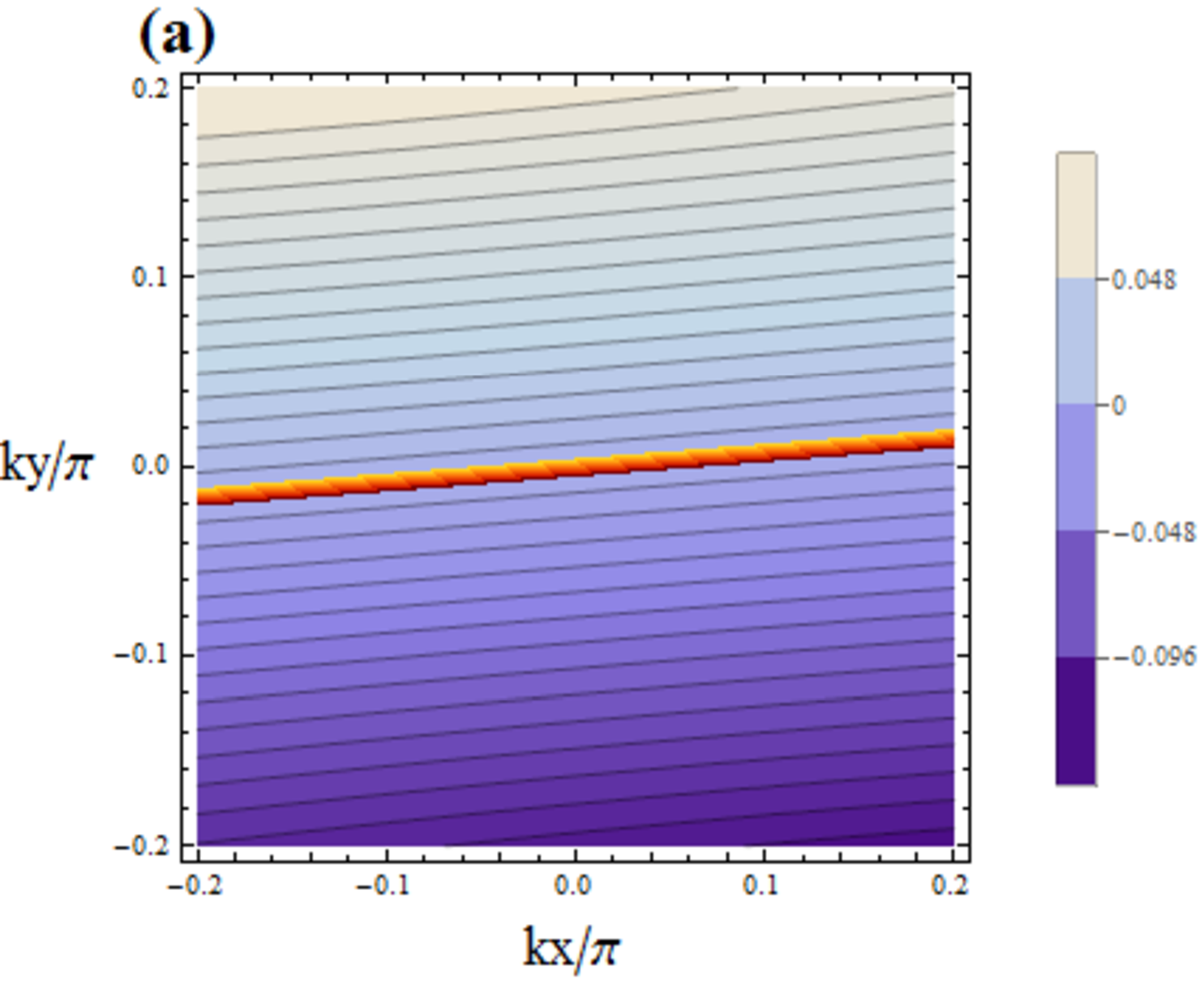}   
\includegraphics[width=4cm]{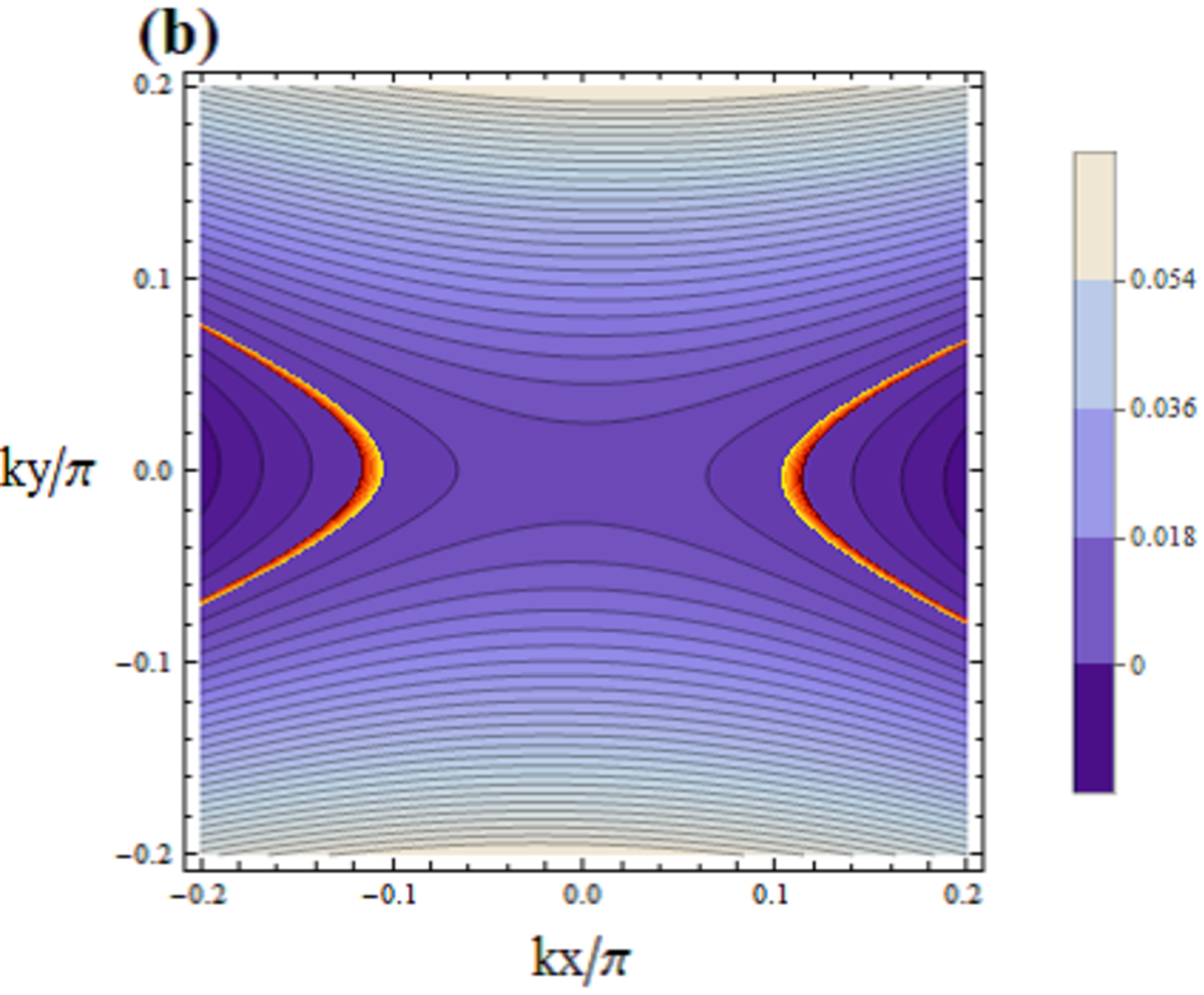}   
\includegraphics[width=4cm]{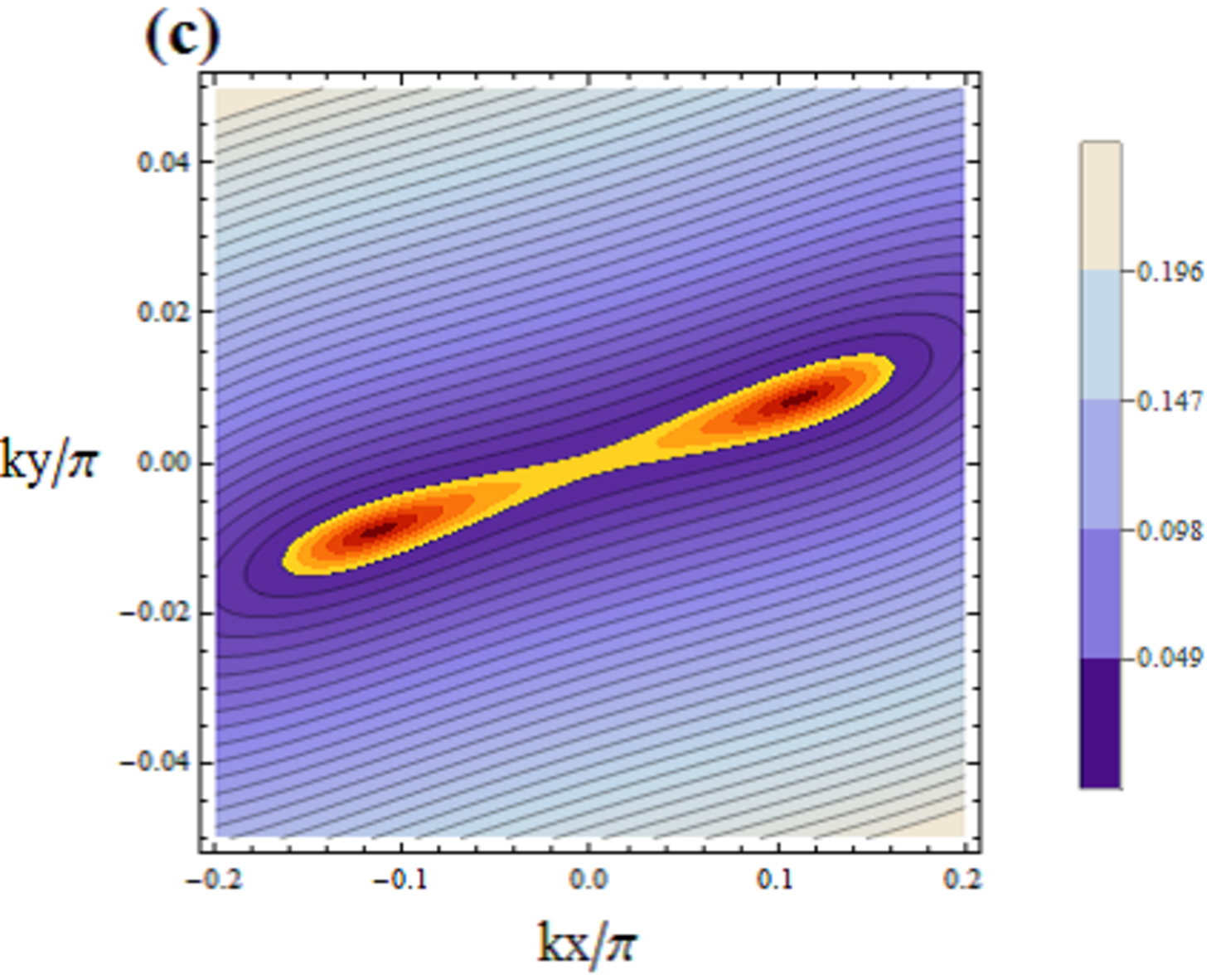}   
\includegraphics[width=3.5cm]{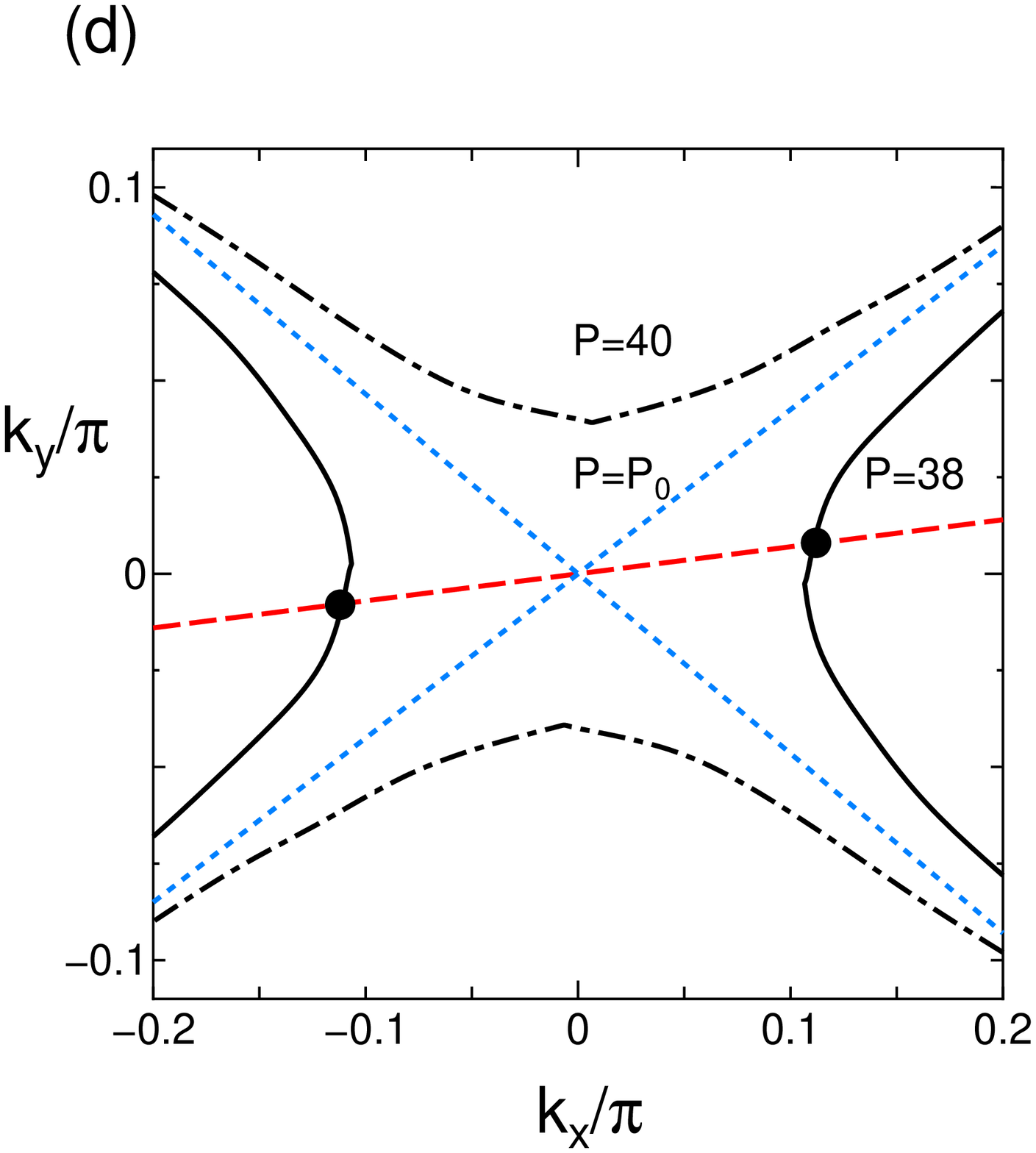} 
  \caption{(Color online)
Contours of  $f(\bm{k})$ (a) and $g(\bm{k})$ (b) 
at $P$ = 38 kbar where the center $\bm{k}=0$ denotes the  $\Gamma$ point. 
 The bright line shows  $f(\bm{k})=0$  and  $g(\bm{k})=0$.
With respect to the  $\Gamma$ point, 
 $f(\bm{k})$ is antisymmetric 
 and  $g(\bm{k})$ is symmetric. 
The corresponding gap function of $\Dk$ (c)  
 shows  a pair of Dirac points (in the bright region) just before the merging. 
In (d), the solid (dot-dashed) line denotes  $g(\bmk)=0$  
 at  $P$= 38 kar (40 kbar) and  
 the dashed line denotes  $f(\bmk)= 0$  for $P \simeq P_0$.
The dotted line denotes  $g(\bm{k})=0$  at $P=P_0$ (= 39.2 kbar).
}
\label{Fig3}
\end{figure}

First,    we show  the merging behavior by taking  Eq.~(\ref{eq:eq12}) up to the first order for simplicity. 
Figure \ref{Fig3}(a) shows $f(\bmk)$ at $P$=38 kbar
 where  $f(\bmk)=0$ (the bright line) gives the  merging direction,  $k_y \simeq 0.07 k_x$. 
Figure \ref{Fig3}(b) shows the corresponding $g(\bmk)$ 
 where the bright line denotes $g(\bmk)=0$, and   $g(0) > 0$ 
at the $\Gamma$ point. 
For $P >P_0$, the bright line moves  from the horizontal axis to the vertical axis  owing to  $g(0) < 0$.
Note that $g(\bmG) <0$ for the X point and  
 $g(\bmG) > 0$ for the Y and M points.
 The gap function $\Dk$ [Eq.~(\ref{eq:eq14})] is obtained  
 in Fig.~\ref{Fig3}(c) representing 
   the contour just before the merging,  
    which is slightly larger than that of the numerical diagonalization.
  The  contour of $\Dk$   close to 
 the Dirac point  is  elliptic and the principal axis is almost parallel 
  to the  merging direction.
The  merging at $\bm{G}/2$  is determined by 
 $g(\bmG)= 0$,  while  $f(\bm{G}/2)=0$ is the identity.  
Figure \ref{Fig3}(d) shows 
 the lines of $g(\bmk)=0$  
 for $P$= 38 kar (solid line)
 and 40 kbar (dot-dashed line), 
 and $f(\bmk)= 0$  for $P \simeq P_0$ (dashed line).
The dotted line denotes $g(\bmk)=0$  at $P=P_0$,  which is given by 
  $k_y \simeq ( -0.02 \pm 0.045) k_x$. 
 The Dirac point exists only for $P <P_0$ 
   owing to the presence of an intersection between $f(\bm{k})=0$ and $g(\bm{k})=0$, which is shown by 
 the closed circle with $(k_x/\pi,k_y/\pi) \simeq \pm (0.112, 0.008)$.
The absence of the intersection for $P_0 < P$ leads to 
the formation of the gap in $\Dk$.
 Figure \ref{Fig3}(d) is also understood by the following analysis with 
  $\bmG = 0$  (the $\Gamma$ point). 
For the pressure  close to   $P=P_0$, 
 Eqs.~(\ref{eq:eq15}) and (\ref{eq:eq16})
  are expanded with respect to $\bm{q} (= \bm{k} - \bmG)$ as 
\begin{eqnarray}
 g(\bm{k}) & = & D(P) - C_{2x}q_x^2 + 2C_{2xy}q_xq_y + C_{2y}q_y^2\; , 
 \label{eq:eq18}
\end{eqnarray} 
 and  $f(\bm{k})  \simeq    - C_{1x}q_x + C_{1y}q_y $, 
 where    
  $D(P) \simeq D_0 (P_0 -P )$.
Figure \ref{Fig3}(d) is well described by the   parameters 
  estimated   as   
$D_0 \simeq 0.0039$, 
 $C_{2x} \simeq 0.386$,  
 $C_{2xy} \simeq 0.040$, and   
 $C_{2y} \simeq 1.95$,  
  $C_{1x} \simeq 0.04$, and   $C_{1y} \simeq 0.56$.

\begin{figure}
  \centering
\includegraphics[width=4cm]{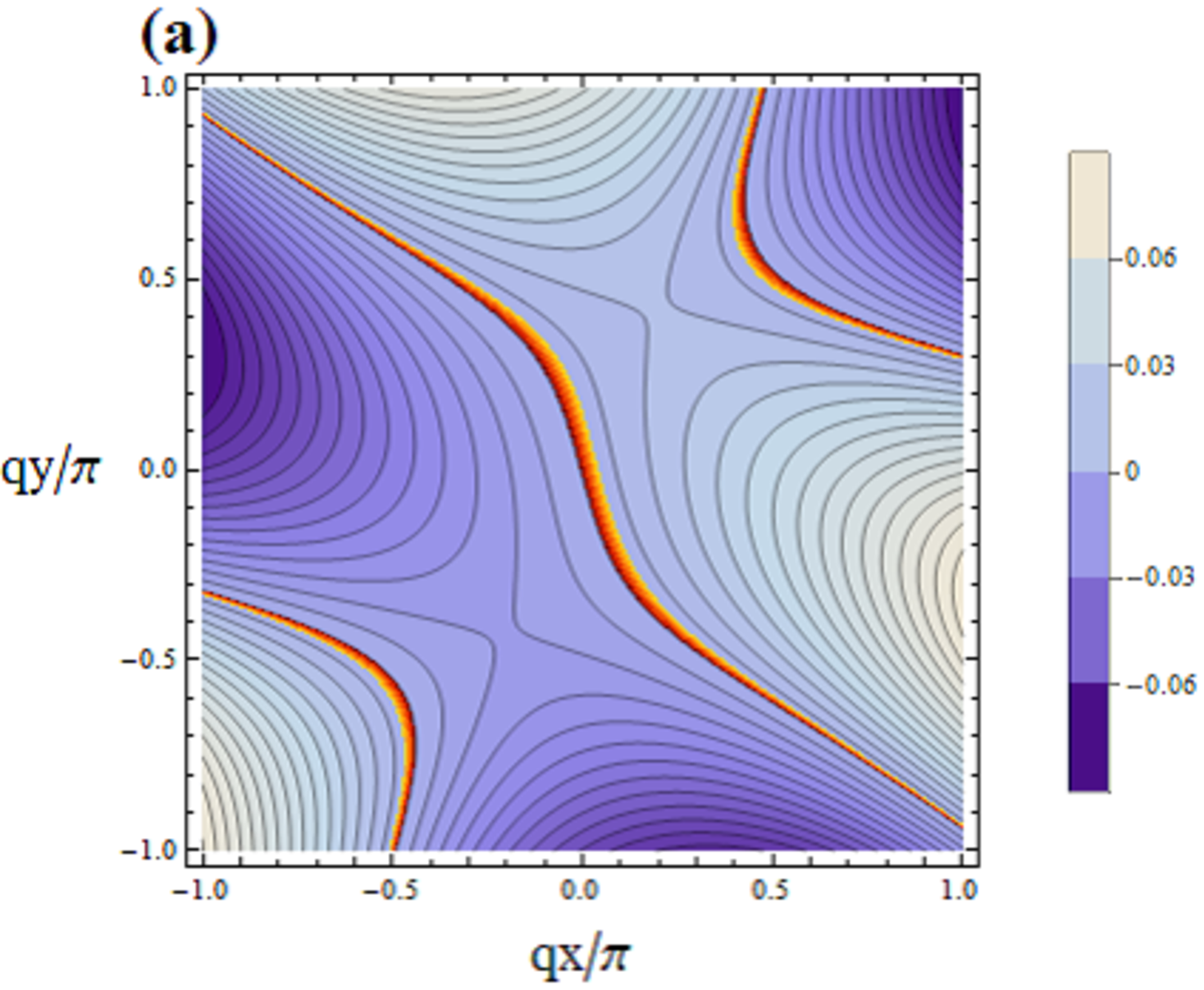}   
\includegraphics[width=4cm]{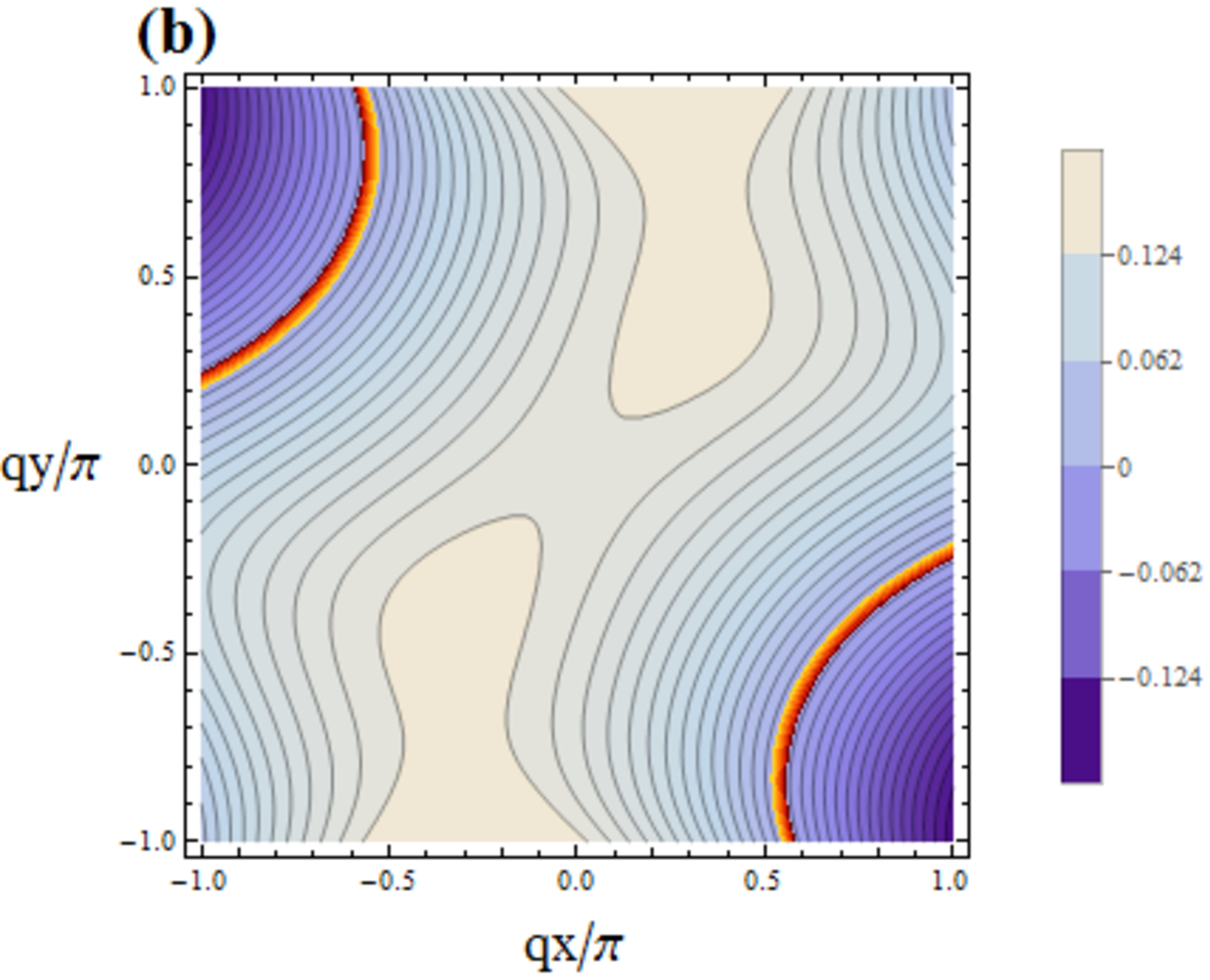}   
\includegraphics[width=4cm]{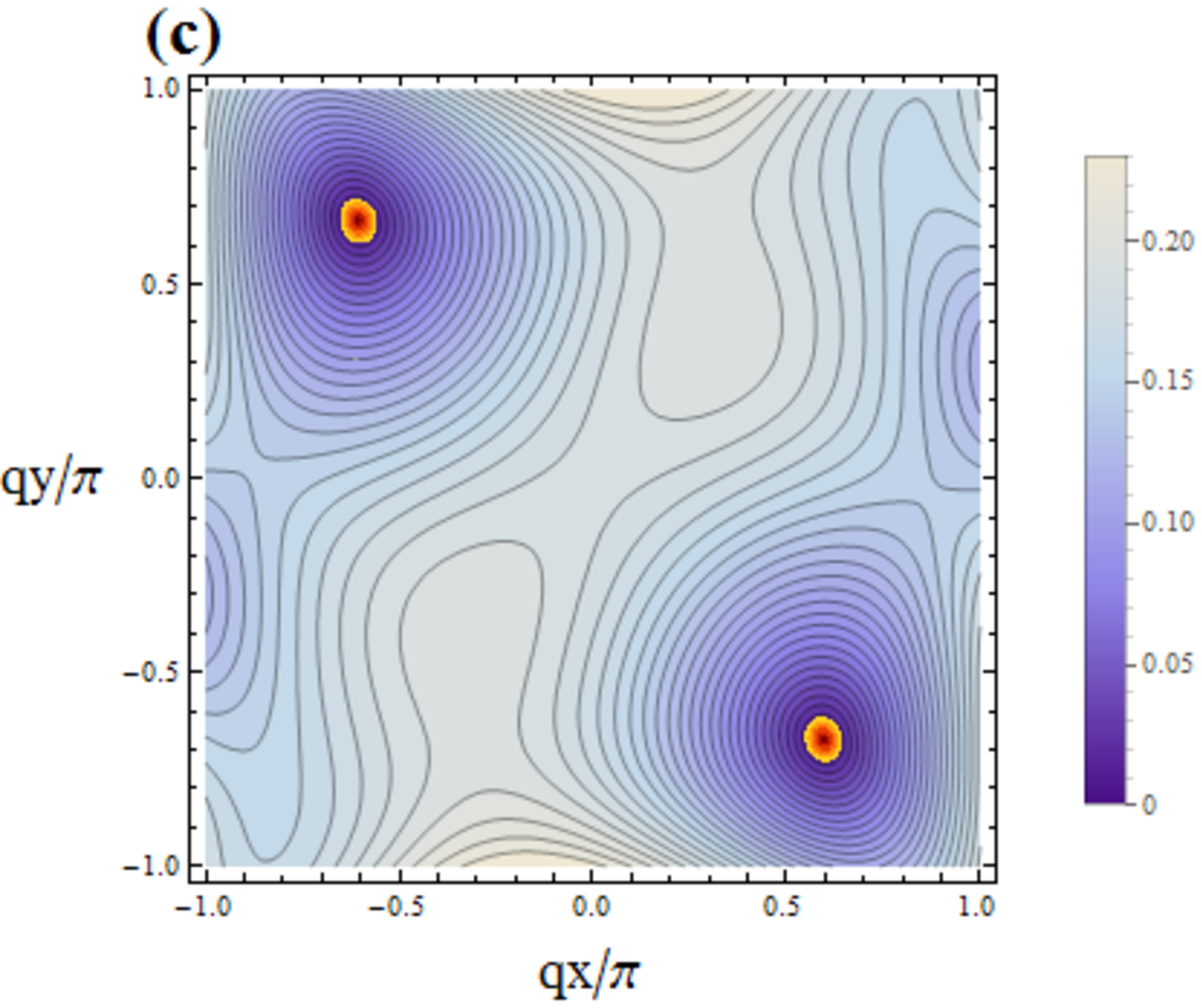}   
\includegraphics[width=4cm]{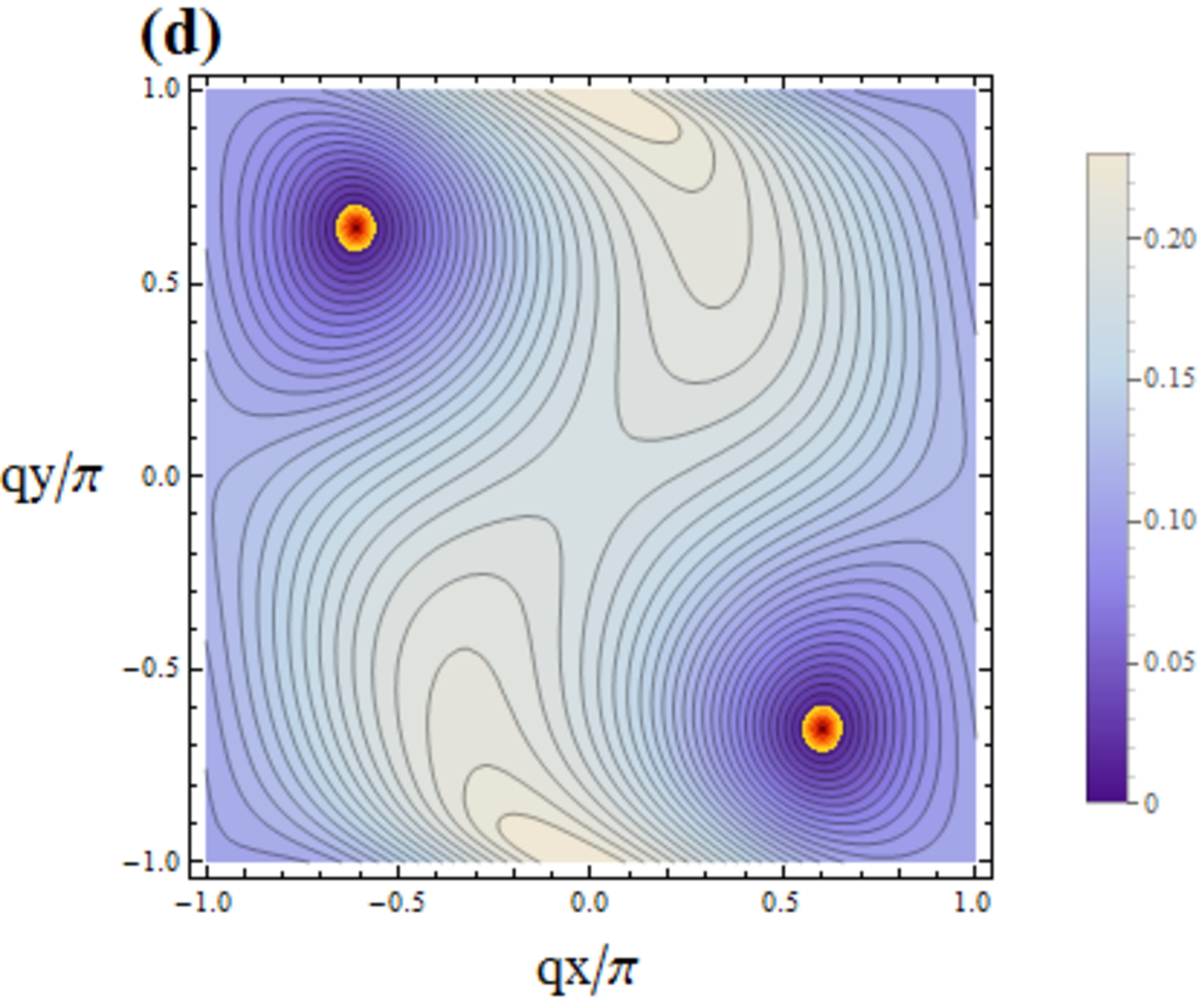}   
  \caption{(Color online)
 $\bm{q} [= \bmk -(0, \pi)]$ dependence of 
    the contour for $f(\bm{q})$ (a),  $g(\bm{q})$ (b),  and 
       $\Delta_{\bm{q}}$ (c) at $P$=4 kbar, 
  which are obtained from Eqs.~(\ref{eq:eq15}), (\ref{eq:eq16}), 
    and (\ref{eq:eq14}) 
       with $\bm{G}/2= (0, \pi)$ (Y point).
The center ($ \bm{q}= 0) $ denotes  the Y point  where 
  $f(\bm{q})$ [$g(\bm{q})$] is antisymmetric (symmetric) with respect to
    $\bm{q}$.
These (a) and (b) give $\Dq$ in (c) 
 and are compared with $\Dq$ in (d),  which is obtained  from the numerical  
  diagonalization of $H(\bmk)$. 
 The Dirac point in the bright region 
 is given by 
 $(q_x/\pi,q_y/\pi) = \pm (0.58, - 0.67)$ for (c) 
 and   $ \pm (0.60, - 0.68)$ for (d).
}
\label{Fig5}
\end{figure}

Next, we examine the Dirac point by applying   Eq.~(\ref{eq:eq12}) up to the second-order perturbation.
 With decreasing pressure, 
  the difference in $\bm{k}_0$
 between the first- and  second-order perturbations increases, and  
 the choice of the $\Gamma$ point  becomes invalid at low pressure, e.g., 
the  obtained Dirac point at $P$=10 kbar exhibits a different behavior. 
Thus, the Dirac point for lower pressures is calculated by taking the Y point 
 where  the matrix element with the antisymmetric function 
  of  $\bm{q} [=\bmk -(0, \pi)]$ is given by 
$\h_{12}$, $\h_{14}$, $\h_{32}$, and $\h_{34}$.
Figure \ref{Fig5}(a) shows $f(\bm{q})$ for $P$= 4 kbar corresponding to 
 the pressure  of the Dirac electron found in \ET.\cite{Kajita_JPSJ2014} 
The bright line obtained from $f(\bm{q})=0$ 
   exists also for the second-order perturbation 
 since the intermediate state $|n>$ with $n$ = 3 and 4 in Eq.~(\ref{eq:eq12}) 
 is orthogonal to  either $|i>$ or $|j>$ with a different parity. 
Note that such a line exists  in the presence and  absence of 
the Dirac point,  
 suggesting an intrinsic  property that originates from 
  the states $|i>$ and $|j>$ with a different parity. 
Figure \ref{Fig5}(b) shows the corresponding $g(\bm{q})$ where $g(\bm{q})=0$ 
 (the bright line) intersects  with $f(\bm{q})=0$ in Fig.~\ref{Fig5}(a). 
The existence of the line of $g(\bmq)=0$ is understood as follows.  
At the Y point (center),  $g(\bmq)>0$  from the definition, and 
  $g(\bmq) < 0$ at the X point  since  
 $<1|\tilde{H}_\bmk|1>$ becomes  smaller than  $<2|\tilde{H}_\bmk|2>$ 
  at the X point,  as seen from  the comparison of each component 
 of the Y and X points in 
 the inset of Fig.~\ref{Fig2}(b).
Figure \ref{Fig5}(c) shows $\Dq$ [Eq.~(\ref{eq:eq14})], 
  which is calculated from
 Figs.~\ref{Fig5}(a) and 4(b). 
 Figure \ref{Fig5}(d) shows $E_1(\bm{q})-E_2(\bm{q})$, which is  numerically 
 calculated by the diagonalization of Eq.~(\ref{eq:eq2}). 
The behavior around the Dirac point shows 
 a good correspondence  between Figs.~\ref{Fig5}(c) and \ref{Fig5}(d),  
 suggesting the validity of Eqs.~(\ref{eq:eq11}) and (\ref{eq:eq12}) 
 for finding  the Dirac point.

\begin{figure}
  \centering
\includegraphics[width=6cm]{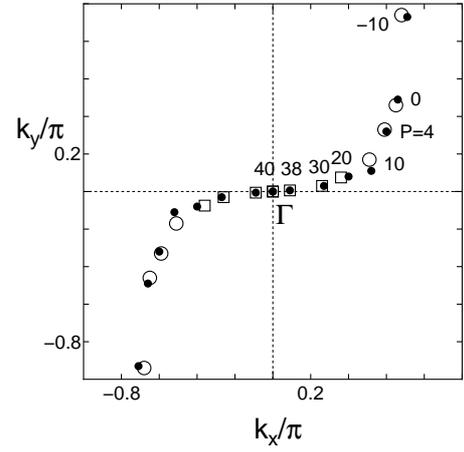}   
  \caption{ 
Dirac point on the plane of 
$k_x$ and $k_y$ with some choices of $P$ (kbar).
 The closed circle is the exact one and the 
 open square (open circle) is obtained  
using the unperturbed state of the $\Gamma$ point (Y point) in  Eq.~(\ref{eq:eq12}).
}
\label{Fig6}
\end{figure}

Figure \ref{Fig6} shows 
 the Dirac point on the plane of 
$k_x$ and $k_y$ with some choices of $P$ (kbar). 
 The closed circle is calculated  by the numerical diagonalization 
 of Eq.~(\ref{eq:eq2}), 
 where the Dirac point moves  from the $\Gamma$ point ($P = P_0$) to the Y point ($P$= -11.8 kbar) with decreasing pressure. 
 The  open square is obtained from  Eq.~(\ref{eq:eq12}) 
  with the  unperturbed state at the $\Gamma$ point. 
 Close to the $\Gamma$ point, the difference between the closed circle and the open square  is small,  
   but increases with decreasing $P$.
 The  open circle is obtained from  Eq.~(\ref{eq:eq12}) 
  with the  unperturbed state at the Y point.
 The difference between the closed circle and the open circle is small 
  but increases with increasing $P$ from 10 kbar. 
As shown in Figs.~\ref{Fig5}(c) and \ref{Fig5}(d), 
the Dirac point and the gap function $\Dq$ at $P$ = 4 kbar are well
 described by Eq.~(\ref{eq:eq11}).
Thus,  the Dirac point is attributable to  
  two ingredients given by  
   Figs.~\ref{Fig5}(a) and \ref{Fig5}(b), which exhibit  
 the antisymmetric  and symmetric functions  with respect to 
 the TRIM (of the Y point) and the resultant lines 
 of $f(\bm{q})=0$ and $g(\bm{q})=0$. 
Finally, we note a  range for using  Eq.~(\ref{eq:eq13})  
 where   the region with  $ 0> \tilde{E}_1(\bmk) - E_1(\bmk)> -0.02$
 covers a wide range  including  the Y and  $\Gamma$ points, 
  and the  Dirac point. 
 However, although the location of the Dirac point is successfully obtained,
  the present method  gives the overtilted Dirac cone for Fig.~\ref{Fig5}(c)  
 owing to the first term of Eq.~(\ref{eq:eq13}),
 and  remains  to be explored as  the next step.

In summary, 
 using an effective Hamiltonian 
 based on   two states with even and odd parities at the TRIM, 
 we obtained the Dirac point as the intersection of two lines. 
One comes from  the diagonal element with the symmetric function 
 relevant to the crossing of two energy bands, 
 and the other originates from the off-diagonal element 
 with the antisymmetric function 
  leading to   the vanishing of the gap.  

%
\acknowledgements
The author thanks R. Kato for useful discussions. 
This work was supported 
 by a Grant-in-Aid for Scientific Research (A)
(No. 15H02108) and (C)  (No. 26400355)
 from the Ministry of Education, Culture, Sports, Science, and Technology, Japan,




\begin{thebibliography}{}
\bibitem{Katayama2006_JPSJ75} 
S. Katayama, A. Kobayashi, and Y. Suzumura, J. Phys. Soc. Jpn.
 \textbf{75}, 054705  (2006).

\bibitem{Kajita_JPSJ2014} 
K. Kajita, Y. Nishio, N. Tajima, Y. Suzumura, and A. Kobayashi, 
 J. Phys. Soc. Jpn. \textbf{83},  072002 (2014).

\bibitem{Novoselov2005_Nature438}
K. S. Novoselov, A. K. Geim, S. V. Morozov, D. Jiang, M. I. Katsnelson,  
I. V. Grigorieva, S. V. Dubonos, and A. A. Firsov,
Nature {\bf 438}, 197 (2005).

\bibitem{Ando2005_JPSJ74} 
For example, see  the review by T. Ando, J. Phys. Soc. Jpn {\bf 74},  777 (2005).

\bibitem{Herring}
C. Herring, Phys. Rev. {\bf 52},  365 (1937).


\bibitem{Fu2007_PRB76} 
L. Fu and C. L. Kane,
Phys. Rev. B {\bf 76},  045302 (2007).

\bibitem{Mori2013_JPSJ} 
T. Mori, J. Phys. Soc. Jpn. \textbf{82},  034712 (2013).

\bibitem{Piechon2013_JPSJ} 
F. Pi\'echon and Y. Suzumura,
 J. Phys. Soc. Jpn. \textbf{82},  033703 (2013).


\bibitem{Kariyado2013_PRB} 
T. Kariyado and Y. Hatsugai, 
Phys. Rev. B {\bf 88}, 245126 (2013).


\bibitem{Kobayashi2013_JPSJ} 
A. Kobayashi and Y. Suzumura, J. Phys. Soc. Jpn. \textbf{82}, 054715  (2013).

\bibitem{Piechon2013_Berry} 
F. Pi\'echon and Y. Suzumura, 
 J. Phys. Soc. Jpn. \textbf{82},  123703 (2013).

\bibitem{Suzumura2013_JPSJ} 
Y. Suzumura, T. Morinari, and F. Pi\'echon,
  J. Phys. Soc. Jpn. \textbf{82},  023708 (2013).


\bibitem{Mori1984_CL} 
T. Mori, A. Kobayashi, T. Sasaki, H. Kobayashi, G. Saito, 
 and H. Inokuchi, Chem. Lett. {\bf 13}, 957 (1984). 


\bibitem{Katayama2009_EPJB57}
S. Katayama, A. Kobayashi, and Y. Suzumura, 
Eur. Phys. J. B. {\bf 67},  139 (2009).


\bibitem{Kobayashi2010_PRB84} 
A. Kobayashi, Y. Suzumura,  F. Pi$\acute{\rm e}$chon, 
 and  G. Montambaux, Phys. Rev. B {\bf 84},  075450 (2011). 


\bibitem{Kariyado2012} 
T. Kariyado and M. Ogata, J. Phys. Soc. Jpn. \textbf{80}, 083704  (2011); 
T. Kariyado and M. Ogata, J. Phys. Soc. Jpn. \textbf{81},  064701 (2012).

\bibitem{Kondo_2005} 
R. Kondo, S. Kagoshima, and J. Harada, Rev. Sci. Instrum. {\bf 76}, 
 093902 (2005).

\end{thebibliography}
\end{document}